\documentclass[twocolumn, twocolappendix, times]{aastex631}
\usepackage{float}
\usepackage{comment}
\usepackage{xcolor}
\usepackage{soul}
\usepackage{amsmath}
\hypersetup{linkcolor=blue,citecolor=blue,filecolor=cyan,urlcolor=magenta}

\newcommand{\msun}{M$_\odot$ }
\newcommand{\psqcm}{cm$^{-2}$ }

\newcommand{\ciii}{\ion{C}{3} }
\newcommand{\civ}{\ion{C}{4} }
\newcommand{\nv}{\ion{N}{5} }
\newcommand{\siiv}{\ion{Si}{4} }
\newcommand{\ovi}{\ion{O}{6} }
\newcommand{\ovii}{\ion{O}{7} }

\setstcolor{cyan}

\setstcolor{red}

\shorttitle{Warm Ions in IGM}
\shortauthors{Bromberg \textit{et. al.}}

\begin{document}

\title{Intergalactic-Absorption Confounding Circumgalactic Observations}
\correspondingauthor{Itai Bromberg}
\email{itai.bromberg@mail.huji.ac.il}

\author{Itai Bromberg}
\affiliation{Centre for Astrophysics and Planetary Science, Racah Institute of Physics, The Hebrew University, Jerusalem 91904, Israel}

\author{Kartick C. Sarkar}
\affiliation{Dept of Astrophysics, Tel Aviv University, Tel Aviv, Israel}
\affiliation{Dept of Space, Planetary \& Astronomical Sciences and Engineering, Indian Institute of Technology Kanpur, Kanpur, India}

\author{Orly Gnat}
\affiliation{Centre for Astrophysics and Planetary Science, Racah Institute of Physics, The Hebrew University, Jerusalem 91904, Israel}

\author{Yuval Birnboim}
\affiliation{Centre for Astrophysics and Planetary Science, Racah Institute of Physics, The Hebrew University, Jerusalem 91904, Israel}

\begin{abstract}
The origin of warm ions in the circum-galactic medium (CGM) surrounding massive galaxies remains a mystery. 
In this paper, we argue that a significant fraction of the observed warm-ion columns may arise in the intergalactic medium (IGM) surrounding galactic halos. 
We use a simple spherical collapse model of the dark matter (DM) halos and their baryonic content to compute the evolving ion fractions within and outside virial halos. 
We show that the photoionized IGM may produce a thick blanket of warm ions around the CGM, thereby contaminating CGM observations. 
We find that the IGM contributes $> 75\%$ of the total \ovi column densities in halos with virial masses exceeding a few times $10^{11}~M_\odot$, and that it may dominate the \ovi absorption even for lower mass-halos, depending on the impact parameter.
We compare our results with observations and find that our simplified model reproduces the overall \ovi columns as well as their trend with the impact parameter and halo mass. 
We show that observed warm ion columns may be completely dominated by the IGM envelopes, consistent with CGM$^2$ data. We, therefore, suggest that theoretical interpretations of CGM-survey observations must consider the possible contribution of the surrounding IGM.
Although our simplified model suggests that it may be possible to kinematically distinguish between CGM and IGM origins through the absorption line profiles, this distinction is likely unfeasible in realistic astrophysical halos, due to the complex velocity structure in the multi-phased CGM.
\end{abstract}
\keywords{galaxies: general, halos, intergalactic medium -- ultraviolet: galaxies}

\section{Introduction} \label{sec:intro}

The circum-galactic medium (CGM) plays an essential role in galaxy formation and evolution, by controlling gas inflow into galaxies and outflow from them. 
While various aspects of the complex, multi-phased, CGM have been extensively studied, the roles that different physical processes play in determining its internal structure and ionization properties remain unclear and are the focus of intense observational and theoretical work \citep{Prochaska2011, Tumlinson2011, Shen2012, Muzahid2012, Werk2013, Lehner2014, Werk_2016, Oppenheimer2016, Stern2016, Gutcke2017, Suresh2017, Tumlinson2017, Faerman_2017, Faerman_2020}. 
The cosmological theory of galaxy formation indicates that galactic halos form as gas from the diffuse intergalactic medium (IGM) falls onto the dense regions where galaxies form. While dark matter (DM) particles oscillate about the center of the potential well, baryons are shock-heated to a ``virial temperature", corresponding to the mass of the forming galaxy. These shock-heated halos are natural sites for the production of highly ionized species. However, warm ions may also be produced outside the virial radius, in the surrounding intergalactic medium. In this region, the baryonic density and temperature profiles are still affected by the halo's gravity, as well as by radiation fields (shock-self radiation, $UV$ background, etc), strong outflows, cosmic rays, and the inhomogeneity of the cosmic web.

Observational campaigns such as the \textit{COS}-Halos survey \citep{Tumlinson2011}, IMACS survey \citep{Johnson_2015}, or CGM$^2$ survey \citep{Tchernyshyov2022a} have examined the extent of warm ions (e.g. \civ, \ovi) surrounding galaxies, as revealed via absorption features in the spectra of background-quasars. These studies focus on $\gtrsim L_\star$ galaxies ($10^{11.5}\lesssim M_{\rm halo}\lesssim10^{13}~M_{\odot}$) at various impact parameters ($\rm 10\lesssim b \lesssim400~\rm kpc$) and relatively low redshifts ($0.1\lesssim z \lesssim0.6$).
The \textit{COS}-Halos survey \citep{Tumlinson2011} demonstrates the existence of \ovi absorption, at impact parameters up to $150$ kpc, with no apparent trend of the column density with impact-parameter within this range. Later, \cite{Johnson_2015} used a larger data sample (\textit{COS}-Halos+IMACS), and showed that no \ovi absorption is detected at impact parameters above the virial radius,  with upper-limits $\sim 10^{13.4}$ \psqcm. Their findings suggest that \ovi resides primarily within the halo\footnote{However, they state that upper limits of $N_{OVI}>10^{14}$ \psqcm are not shown, due to constraints sensitivity considerations. See \cite{Johnson_2015} for details}.
\cite{Tchernyshyov2022a} used the CGM$^2$ data sample to study the extent of \ovi absorption. They found that \ovi is detected out to their maximal surveyed impact parameter, of $\sim 400$ kpc, without any particular trend, and with upper limits of order $10^{14}$ \psqcm. According to their fit, significant \ovi columns may persist up to $(2-3) R_{200}$, depending on the halo mass. 
For lower mass ($\gtrsim 0.1L_\star$) galaxies, \cite{Prochaska2011} used the LCO/WFCCD survey \citep{Prochaska_2011_IV} to explore a much larger range of impact parameters, extending out to  
$\sim1000$~kpc.
They found a large covering fraction of \ovi outside the galactic halos, particularly for sub-$L_\star$ galaxies. They associated it with the diffuse medium surrounding individual galaxies.
The prevalence of \ovi in and around galactic halos thus remains inconclusive.

Detailed $3D$ cosmological simulations of galaxy formation attempt to reproduce key trends in observational surveys, by invoking various physical processes and assumptions (e.g. radiative cooling,  ionization, star formation and evolution, chemical enrichment, wind feedback, black hole growth, AGN feedback, cosmic dust, etc.).
\cite{Nelson2018} used the IllustrisTNG simulations, to estimate the \ovi abundance around galaxies. They found that \ovi columns persist out to an impact parameter of $10$ Mpc, due to contributions by nearby satellite halos and the IGM (Figure 9 therein).  They showed that although the halo's contribution to the \ovi column density drops at roughly the virial radius, there are significant contributions from other sources outside the central halo, that prevent a distinct cut-off in the column density. 

\cite{Oppenheimer2018} used the EAGLE zoom-in simulations, to probe \ovi columns at impact parameters below $\sim300$ kpc. Their results indicate a slow decline in the \ovi column with impact parameter, regardless of halo mass (see Figure 2 therein). 
In a later paper by \cite{Ho2021}, an EAGLE simulation also produced a flat \ovi column density vs. impact parameter profile, with contributions from gas outside $3 R_{\rm vir}$.
\cite{Appelby2022} used SIMBA to study different ion absorption features and their origin. They estimated that \ovi absorption is comprised of roughly equal contributions from the gas inside and outside the halo (their Figure 7, and analysis therein).

An alternative to cosmological hydrodynamic simulations is the use of semi-analytic $1D$ models, which aim to explain the observed trends by focusing on key physical processes. 
Several works have focused on identifying the origin of \ion{O}{6}, and the ionization mechanisms which dominated its production.
\cite{Stern_2018} considered the possibility that \ovi is produced either in the shocked halo or in a low-pressure photo-ionized layer outside the accretion shock. Their low-pressure scenario requires that the accretion shock be located well within $R_{\rm vir}$. This low-pressure \ovi phase is disfavored by cosmological simulations but can reproduce the observed properties of \ovi as well as other absorption features. 

\cite{McQuinn_2018} and \cite{Qu_2018} explored the dominant ionization mechanism for \ovi production in the CGM.  \cite{McQuinn_2018} included radiative cooling and feedback and showed that inside the halo, \ovi is probably produced collisionally, with large cooling gas flows regulated by feedback.  \cite{Qu_2018} considered a similar model but also took into account photoionization and steady-state cooling, and demonstrated that photoionization may play a key role in lower mass halos, as well as in the outskirts of massive halos, due to the lower densities.

\cite{Voit_2019} introduced precipitation-regulated feedback models, which account for the relatively constant \ovi column density over large spans of halo masses and impact parameters. The emphasis on precipitation has been inspired by the velocity profiles of detected \ovi, which do not exceed the halos' escape velocity. This suggests that the gas is circulated inside the halo. Furthermore, this supports the notion that the CGM is multiphased, consistent with the simultaneous detection of various ionic species.

In \cite{Faerman_2017}, a $1D$ multiphased semi-analytic model based on isothermal hydrostatic profiles has been presented. Assuming turbulent-driven log-normal distributions for the densities and temperatures of both the warm ($\rm \sim10^4K$) and hot phases, the model reproduced the detections and upper limits from \cite{Tumlinson2011} and \cite{Johnson_2015}. \cite{Faerman_2020}  extended the multiphased model to isentropic profiles, and allowed for radius-dependent temperatures and non-thermal contributions. Their profiles for the hot gas were calibrated using data from \cite{Tumlinson2011} and \cite{Johnson_2015}. As these models extend only slightly beyond $R_{\rm{vir}}$, they predict a decline in column density at larger radii.

While warm-ions column densities serve as common diagnostics for CGM analysis and interpretation, theoretical models do not usually account for the warm ions produced outside the shock radius, where the halo's presence still affects the baryon-density profile. Computationally, even though simulation-based studies may naturally include this component, they do not focus on the role that the IGM plays in absorption line observations. This is particularly true for low mass galaxies ($M_{\rm vir} \lesssim 10^{11}$ \msun) where the observations show a significant presence of \ovi in contrast to the theoretical CGM models that fail to predict such behavior. This further motivates us to study the IGM contribution to the warm ions at different mass scales. As we will see, in our model the origin of \ovi around low-mass galaxies is mainly from the surrounding IGM and not the CGM. 

In this work, we argue that the gravitational focusing around halos, which is a natural and unavoidable consequence of structure formation in $\Lambda$CDM cosmology, creates a favorable site for warm-ion production (including \ion{O}{6}) via photoionization by the ambient background radiation. We demonstrate this notion using a highly simplified $1D$ toy model. We show that significant amounts of \ovi should naturally arise outside $R_{\rm shock}$, thus increasing $-$ and sometimes dominating $-$ the observed columns of \ovi and other key ions around galaxies.

In \S\ref{sec_model} we describe our simplified $1D$ model. This includes the evolving properties of the dark matter halos and the distribution and ionization properties of the baryonic matter both within and outside galactic halos. In \S\ref{sec_results} we discuss the observational signatures that arise in our simplified halos. We compute the expected absorption line columns and discuss the relative contributions of the shocked gas within the halo versus photoionized gas in the IGM to the total absorption-line signatures. We discuss the possibility of discerning them observationally based on their kinematic properties. We conclude in \S\ref{sec_conclusions}.

\section{Model description}
\label{sec_model}

We use a simple $1D$ cosmological toy model to study the impact of the halos' gravitational potentials and the $UV$ background on the warm-ions production in halos and their surrounding IGM. The model consists of three major components - i) A redshift-dependent dark matter density profile, ii) Evolving baryonic density and velocity profiles, and iii) The corresponding temperature and ionization states of the baryonic matter. We describe these model ingredients below.

\subsection{Spherical Collapse Model}

A simplified model for the dark matter halo profiles is constructed by tracing the evolution of initial overdensity perturbations solely under the influence of gravity, in an Einstein-de-Sitter (EdS) cosmology. We follow the numerical procedures used in \citet{Birnboim2003} to set the initial conditions for each $1D$ spherical simulation. Our profiles are constructed from shells, each tracing a cycloid of the form,
\begin{equation}\label{eq1}
r=A(1-\cos\eta)
\end{equation}
\begin{equation}\label{eq2}
t=B(\eta-\sin\eta)
\end{equation}
\citep[e.g.][]{ryden2017introduction}, where $\eta$ is the conformal time.

We determine each shell's free parameters ($A$ and $B$) by imposing two requirements: one at an early linear stage, and the other at the epoch of virialization.
First, we require that the mass enclosed within $r$ at some early linear stage ($\eta\ll 1$), satisfies that the overdensity at that time, $\delta(M)\equiv \bar\rho / \rho_u -1$ is proportional to the correlation function of the power spectrum of the universe \citep{Dekel_1981}. In this expression, $\bar\rho$ is the average density inside $r$, and $\rho_u$ is the universal density.
The proportionality constant between $\delta(M)$ and the correlation function can be determined by requiring that a specific mass virializes at a specific time after the Big Bang, within our EdS framework. 
For each shell, we fix $B$ by requiring that at the time of the shell's perspective virialization (when $\eta=3\pi/2$), the radius $r$ and the time $t$ will satisfy $r=r_{\rm vir}$ and $t=r_{\rm vir}/v_{\rm vir}$, with $v_{\rm vir}^2=GM/r_{\rm vir}$ with $M$ the mass enclosed within that shell. This ensures that the infall velocity at this stage is $v_{\rm vir}$. 

Here, we calculate profiles with collapsed masses (at $z=0$) between $10^{11}-10^{14}~M_{\odot}$. We solve eqns. \ref{eq1} and \ref{eq2} for $1000$ shells with the outermost shells exceeding the virial and turnaround radii. The dark matter density at a given radius is calculated numerically by differentiating the mass with respect to the radii of two adjacent shells at each time  \citep[for more details see][]{Birnboim2003}, and the results are then mapped onto a regular $r-t$ grid extending to $10$ Mpc.

\subsection{Gaseous Halo}
Given the evolving dark matter profiles, we compute the properties of the baryonic matter both within and outside the shock radius, defined as:

\begin{equation}
R_{\rm shock}=f_{\rm shock} R_{\rm vir},
\end{equation}
where $f_{\rm shock}$ is a dimensionless multiplication factor.
We explore different values of $f_{\rm shock}$ ($0.5, 1, 1.5,$ and $2$) to account for uncertainties in the extent of the virialized gas. The virial radii of halos with present-day DM masses of $10^{11},10^{12},10^{13},10^{14}~M_{\odot} $, at redshift $z=0.2$, are $R_{\rm vir}\approx80,170, 365,790$ kpc, respectively.

We set the gas density to be everywhere proportional to the evolving dark matter density,
\begin{equation}
    \rho_{\rm gas}(r) = f_b \rho_{\rm dm}(r)\,,
    \label{eq:rho_gas}
\end{equation}
with $f_b = 0.165$, the universal baryonic fraction \citep{Shull_2012, Planck_Collaboration_2016} \citep[c.f. $f_b = 0.13$ in][]{Birnboim2003}.

For the velocities, we follow \citet{Birnboim2003} and set
\begin{equation}
    v=v_{\rm vir} \frac{\sin\eta}{1-\cos\eta}
\end{equation}
outside $R_{\rm shock}$. Inside the halo, we set $v_{\rm CGM}=0$.

\subsection{Metallicity}

For the baryonic component, we assume a metallicity,
\begin{equation*}
  Z=\begin{cases}
    0.03~Z_{\odot}, & \text{ $r\geq R_{\rm shock}$}\\
    0.1~Z_{\odot}, & \text{ $r < R_{\rm shock}$},
  \end{cases}
\end{equation*}
inspired by CGM \citep[e.g.][]{Lehner_2019} and IGM \citep[e.g.][$Z=0.02-0.05~Z_\odot$]{Dalton_2021,Dalton_2022} observations. 

Determining the metallicity in and around halos is an active area of research  
\citep[e.g.][]{Wiersma_2011,Gatuzz_2024,Strawn_2024}, with a broad range of possible values. Consequently, the absorption line column densities may need to be adjusted once a "typical" metallicity is established.
In the CGM, the metal-ion columns are generally proportional to the metallicity. The IGM columns are slightly more sensitive to metallicity due to the impact it may have on the thermal properties of the gas (see below).

\subsection{Temperature and Ionization State}

When determining the thermal and ionization properties of the gas, we assume that the gas is everywhere (both within and outside the halo) exposed to the \cite{Haardt2012} redshift-dependent metagalactic $UV$ background radiation field.
Given the radiation field, we use CLOUDY-v17.01 \citep{ferland2017,ferland2023}, to obtain the thermal equilibrium temperatures, and ion fractions as described below.

We assume that the gas within $R_{\rm shock}$ has a uniform metallicity, $Z=0.1~Z_{\odot}$, and a  temperature $T_{\rm CGM}$ such that 
\begin{equation}
    T_{\rm CGM}=\rm max (T_{\rm PIE },T_{\rm vir}).
    \label{eq:T_cgm}
\end{equation}
In this expression, $T_{\rm vir}$ is the virial (collisional shock) temperature, which depends on the halo's mass and the redshift. We follow  \cite{Johnson_2012} (eq. 1 therein) and set,
\begin{equation}
       T_{\rm vir} = 4 \times 10^4 \left(\frac{\mu}{1.2}\right) \left(\frac{M_{\rm vir}}{10^8 h^{-1} M_\odot}\right)^{2/3} \left(\frac{1+z}{10}\right)~{\rm K},
\end{equation} 
where $h\approx0.7$ is the dimensionless Hubble constant $\mu$ is the mean molecular weight ($\approx 0.6$ for a fully ionized gas).

$T_{\rm PIE}$ is the thermal equilibrium temperature which, given the redshift-dependent background radiation, is a function of the local density, $\rho_{\rm gas}(r)$. 

For the ionization of the CGM gas, we take into account the combined impact of collisional ionization at the (forced) halo temperature $T_{\rm CGM}$ (eqn. 6), and photoionization by the ambient background. We refer to this ionization state as the collisional photo-ionization equilibrium (CPIE).

For the IGM, the ionization and thermal states of the gas are computed assuming a thermal- and ionization equilibrium in the presence of a metagalactic $UV$ background.

\subsection{Column Densities}
Given the run of ion fractions with radius, we can now compute the column densities as functions of the impact parameter. The column density of element $m$ with ionization level $i$ is given by:
\begin{equation}
N_i^m=\int
n_i^m(s) ds= 2\int_{b}^{r_{\rm max}} \frac{n^m(r) x_i^m(r)}{\sqrt{1-\left(\frac{b}{r}\right)^2}} dr
\end{equation}
where we integrate along a line-of-sight intersecting the halo at an impact parameter $b$. $n_i^m(s)$ is the number density of element $m$ in ionization state $i$. 
In the second integral, $x^m_i$ is the ion fraction of element $m$ at ionization level $i$, $n^m=n_{\rm H}\sl A^mZ$ is the number density of element $m$, $n_{\rm H}$ is the total Hydrogen number density, $A^m$ is the solar elemental abundance, and $Z$ is the metallicity. The factor of $2$ is a geometrical factor that accounts for the two sides of the spherical halo. 

\begin{figure*}
\centering
\includegraphics[width=0.9\textwidth]{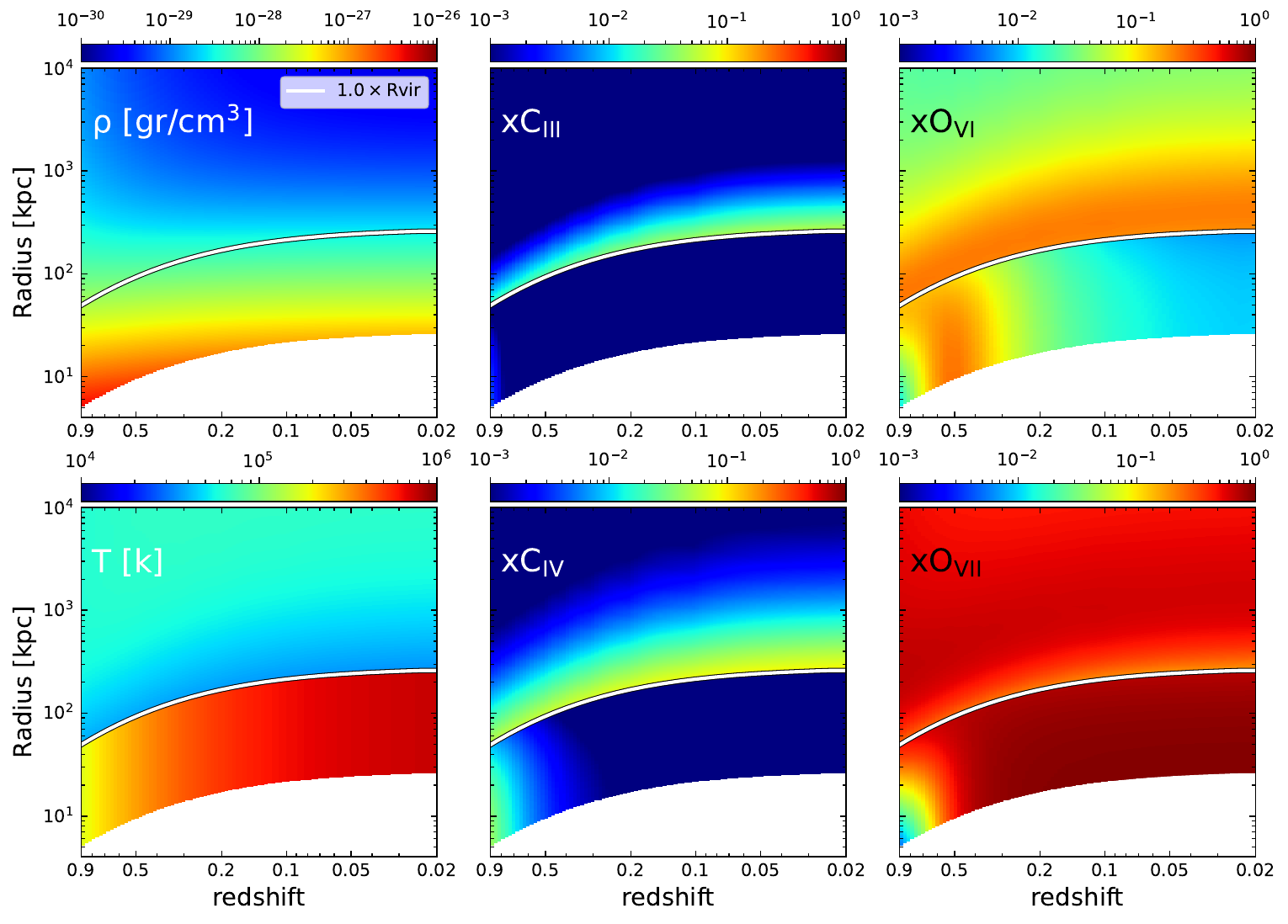}
\caption{Radius-redshift diagrams, for a $M_{z=0} = 10^{12}~M_{\odot}$ halo, color-coded by different quantities. Top-left: baryonic gas density; bottom-left: gas temperature, top middle: \ciii ion fraction, bottom-middle: \civ ion fraction, top-right: \ovi ion fraction, bottom-right: \ovii ion fraction. The white curve in each panel shows the shock radius vs. redshift, with $f_{\rm shock}=1$. We do not consider radii below $0.1 R_{\rm vir}$.}
\label{fig:zr-ions}
\end{figure*}

Since we focus on the ions produced outside the galactic disk, we do not integrate over radii smaller than $0.1 R_{\rm vir}$. The interstellar medium within the galactic disk would highly contaminate any detection at lower impact parameters. 

When estimating the IGM contribution to the total absorption line signatures, we set the maximal integration radius, $r_{\rm max}=3$ Mpc. In our model, this outer radius corresponds to an expansion velocity offset of approximately $\rm \Delta v \simeq 200~km/s$ for halos with masses in the range\footnote{However, for more massive halos, the included velocity range beyond the turnaround radius decreases, whereas for group-sized objects $\gtrsim 5 \times 10^{13}~$M$_\odot$, this radius remains within the collapsing region.} of $ \sim 10^{11} - 10^{12}~$M$_\odot$ at $z\simeq0.2$.

\subsection{Absorption Lines Kinematics}
\label{subsec:line-profile}
Given the density, temperature, and velocity distribution in our halos, we compute the kinematic line profiles, at a given impact parameter. In doing so, we assume that the line-profile within each shell is well approximated by a Gaussian profile at the appropriate temperature, centered about the shells' velocity, and we accumulate the optical depth per frequency through the different shells. The numerical procedure is briefly described in appendix \ref{Appendix: Absorption}.

To validate this kinematic approximation, we also use the \textsc{Trident} Python package \citep{Hummels_2017}. \textsc{Trident} computes absorption line profiles given the density, temperature, and velocity profiles. This module is set to simulate an observation with the \textit{Cosmic Origins Spectrograph} on board the \textit{Hubble Space Telescope}, using a G130M line-spread function kernel. \textsc{Trident} also introduces the Milky Way foreground and a Gaussian noise to the simulated line, for a more direct comparison with observations.

We performed these calculations, including all the shells between ${\rm max}(0.1R_{\rm vir},b)-3$ Mpc for the total observed columns, and again for the IGM $-$ this time including only shells that are outside the halo, between ${\rm max}(R_{\rm shock},b)-3$ Mpc.

\subsection{Limitations}
Our constructed CGM model is highly simplified and does not accurately capture many details of astrophysical halos: It does not account for the morphological, thermal, and dynamical complexities of the multi-phased CGM or the radiative and dynamical properties of the accretion shocks. Nevertheless, as we will demonstrate in the following sections, it approximately reproduces the observed column densities, both within and beyond $R_{\rm vir}$. Our primary objective in this paper is to assess the contribution of the IGM to the warm-ion absorption, and we consider this simplified model to be adequate for this specific purpose.

\section{Results}
\label{sec_results}

\subsection{Ion fractions}
Figure \ref{fig:zr-ions} shows the evolution of our fiducial halo model, for a halo that attains a mass of $10^{12}~M_{\odot}$ at redshift $z=0$, with $f_{\rm shock}=1$ ($R_{\rm shock}\approx260$ kpc at $z=0$). The evolution of various quantities is shown using color maps on the redshift-radius plane. The bottom-left panel shows the evolution of the gas temperature. The gas temperature is determined either by equilibrium with the $UV$ background radiation at the local density, or by the shock temperature for virialized gas $-$ if it is larger than the equilibrium temperature $-$ following equation \ref{eq:T_cgm}. The baryonic gas density is shown in the top-left panel and follows from equation \ref{eq:rho_gas}.

We display the evolving ion fraction distributions for \ion{C}{3}, \ion{C}{4}, \ion{O}{6}, and \ion{O}{7} in the middle and right columns of Figure \ref{fig:zr-ions}.  The ion fractions at CPIE are functions of temperature, density, and redshift. Note, that while the density profile $\rho(r)$ varies slowly with redshift, the virial temperature and the intensity of the $UV$-background radiation are stronger functions of $z$. Consequently, the ion fractions at a given density are also redshift-dependent.

Figure \ref{fig:zr-ions} confirms that far from the halo, the density approaches its undisturbed mean value, resulting in an equilibrium temperature of $T_{\rm PIE} \sim 5\times 10^4$ K. Warm ions are produced in this dilute warm gas due to photoionization.
As the IGM gas falls toward the halo, the gas density ($\rho_{\rm gas}$) rises while the equilibrium temperature ($T_{\rm PIE}$) drops, enhancing the production of warm ions such as \ciii, \civ, etc.
The figure shows that significant fractions of warm ions are produced outside the virial radius, even at low redshifts. Specifically, \ovi is efficiently produced in the low-density IGM down to $z=0$.  We also see that ions such as \ovi can extend nearly a Mpc beyond the CGM forming a widespread envelope of ions around the halo.

Inside the CGM ($r< R_{\rm shock}$), collisional ionization at the virial temperature is generally the dominant ionization process. As the halo evolves, it becomes more massive, and the CGM temperature increases from $\approx 10^5$~K at $z\sim 1$ to $\approx 10^6$~K at $z \approx 0$. Accordingly, the \ovi ion fraction in CGM increases as the halo evolves, peaking at $z\sim 0.5$ before decreasing again. At $z\lesssim 0.5$, the CGM becomes hot enough to support the production of higher ions like \ovii.

\subsection{Column densities}
\label{subsec:column-density}

To test the detectability of the IGM component, we now compute the column densities at various impact parameters. As mentioned above, we define the CGM column as that arising from $r = 0.1R_{\rm vir}$ to $R_{\rm shock}$, and the IGM column as that arising from $r = R_{\rm shock}$ to $3~{\rm Mpc}$. The total column density is the sum of the two.

\begin{figure}
\centering
\includegraphics[width=0.48\textwidth]{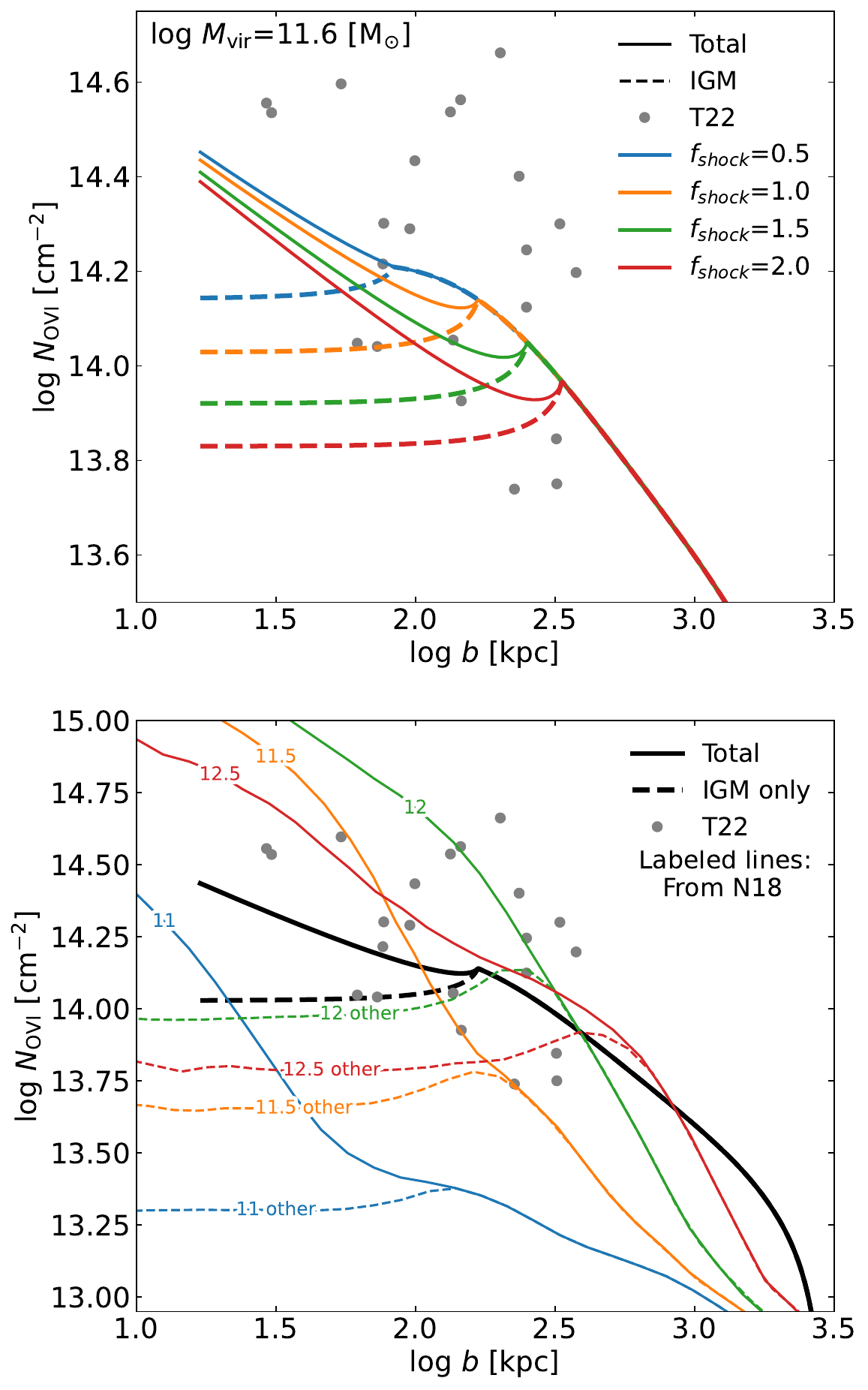}
\caption{Top panel: \ovi column densities vs. impact parameter, for the same halo shown in Figure \ref{fig:zr-ions}, at $z=0.2$ when $M_{\rm vir}(z=0.2)=10^{11.6}~{\rm M}_{\odot}$.
The solid curves show the total column density (from ${\rm max}(0.1 R_{\rm vir},b)$ to $\rm 3~Mpc$). The dashed curves show the IGM contribution (from ${\rm max}(R_{\rm shock},b)$ to $\rm 3~Mpc$). 
Different colors are for different values of $f_{\rm shock}$. 
Bottom panel: A comparison of our model (black curve) for $f_{\rm shock}=1$ with the median \ovi column densities from \cite{Nelson2018} for various halo masses (see labels) at $z=0$. Solid curves are for total column densities, and dashed curves are for the gas beyond the shock radius. In both panels, the gray markers show detections from \cite{Tchernyshyov2022a}.}
\label{fig:columns}
\end{figure}

In the top panel of Figure \ref{fig:columns}, we show the total (solid) and IGM-only (dashed) \ovi column densities, for the same halo shown in Figure \ref{fig:zr-ions}, 
at redshift, $z=0.2$ (when its mass is $10^{11.6}$ \msun). 
The different colors display results for different values of 
$f_{\rm shock}$, vs. impact parameter.
The gray markers show detections from the CGM$^2$ survey, as presented in \cite{Tchernyshyov2022a}. 
The top panel in figure \ref{fig:columns} demonstrates that, regardless of the choice of $f_{\rm shock}$,
our highly simplified model produces \ovi column densities comparable to observations. It also demonstrates that the IGM columns are comparable to both the total column densities and the observations. We thus argue that the near-halo IGM may significantly contribute to observed warm-ion column densities in CGM surveys.

The bottom panel of figure \ref{fig:columns} compares our results for the case of $f_{\rm shock}=1$ (black curves, identical to orange curves in the upper panel), with the median column densities computed by \cite{Nelson2018} from \textit{IllustrisTNG} simulations (colored curves).
In addition to the total column densities, \cite{Nelson2018} also provide 
an ``other" component, emanating from IGM gas and satellite halos.
As in the top panel, solid lines correspond to the total columns and dashed lines are for gas outside the halo. Each color is for a different halo mass (see labels). 
Note, that \cite{Nelson2018} provide columns for $z=0$ halos, while our curve is for a $z=0.2$ halo, to allow for a more direct comparison with observations.
Although our model yields a more gradual slope overall, our IGM component 
is comparable to the $10^{11.5-12.5}~{\rm M}_\odot$
halos ``other" curves by \cite{Nelson2018}. The broad agreement of our computed columns with both observations and complex $3D$-simulations serves as a sanity check when using our simplified model to estimate IGM contributions to \ovi column densities. 

\begin{figure}
\centering
\includegraphics[width=0.45\textwidth]{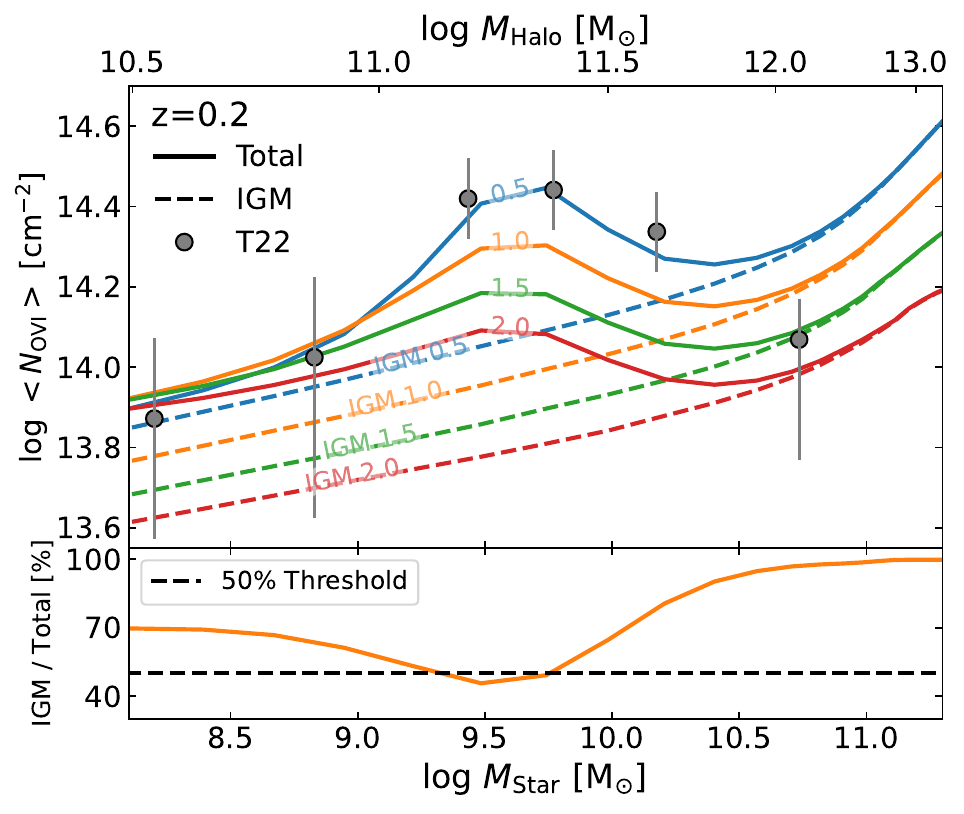}
\caption{Top: Surface-averaged values of \ovi column densities (see text) vs stellar-mass. The solid curves show the total column densities and the dashed curves show the IGM column densities. Different colors are for different values of $f_{\rm shock}$ (see labels).
The gray markers display the CGM$^2$ results from \cite{Tchernyshyov2022a}. Bottom: fraction (percent) of the IGM-\ovi column from the total \ovi column densities, for $f_{\rm shock}=1$. 
The dashed horizontal line marks a $50$\% IGM contribution to the total estimated columns.}
\label{fig:Noviavg_vs_mstar}
\end{figure}

To explore how the overall IGM contribution fractions depend on halo mass, and to compare our results with other works, we now consider surface-averaged columns $-$ representing the entire halo $-$ as was previously done in \cite{Nelson2018} and \cite{Tchernyshyov2022a}. 

To determine the surface averages, we first calculate the total- and IGM-columns\footnote{As before, integrating from ${\rm max}(0.1R_{\rm vir},b)$ to $3$~Mpc for the total column, and from $R_{\rm shock}$ to $3$~Mpc for the IGM columns} as functions of the impact parameter, $b$. Subsequently, we surface-average these columns over impact parameters ranging from $b=0.1R_{\rm vir}$ to $b=R_{\rm shock}$ 
enabling a direct comparison with published results.

The top panel of figure \ref{fig:Noviavg_vs_mstar} shows the surface-averaged total (solid) and IGM-only (dashed) \ovi column densities 
vs. the halo's stellar mass $\rm M_{Star}$. 
The corresponding halo masses are shown on the top axis.
The stellar masses were obtained using the stellar-to-halo mass relation presented in \cite{Girelli2020} (equation 6 and table 1 therein). Different colors are for different values of $f_{\rm shock}$. 
The gray markers display the $\rm CGM^2$ detections from \cite{Tchernyshyov2022a}. We find a good agreement between the observations and our model's total column densities, for $f_{\rm shock}=1$, even when the CGM dominates.
We note that our highly simplified model exhibits a trend similar to the \textit{IllustrisTNG} results presented in \cite{Nelson2018}, albeit shifted by approximately half a decade in mass (see Figure 6 in \cite{Tchernyshyov2022a}). Our model aligns more closely with observational data, whereas the \textit{IllustrisTNG} results tend to favor higher masses by about a factor of $3$.

The nice agreement of our model with the CGM$^2$ data at the lower mass range ($M_{\rm vir} \lesssim 10^{11}~{\rm M}_\odot$) is particularly important: Low-mass galaxies are not expected to host stable CGMs due to the instability of their virial shocks {\citep{Birnboim2003, Dekel2006}}. Therefore, any {\ion{O}{6}} observed around these galaxies must originate from the IGM. The agreement of our basic IGM model with observations in this mass range supports the idea that a simple, photoionized IGM plays a significant role in warm-ion absorption line observations.

\begin{figure*}
\centering
\includegraphics[width=1.0\textwidth, clip=True, trim={0cm 0cm 0cm 2cm}]{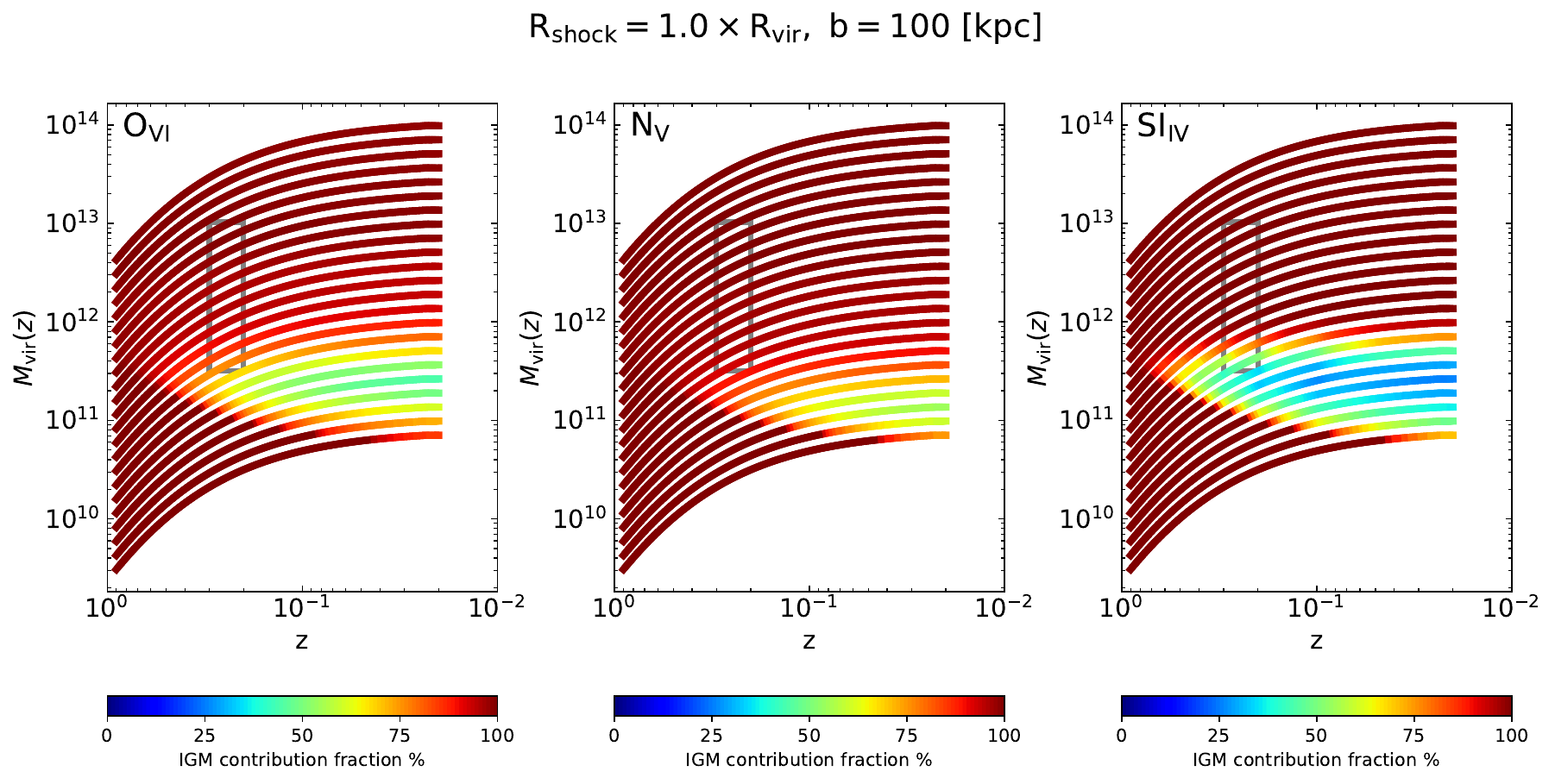}
\caption{Virial mass vs. redshift curves, color-coded by the fraction of IGM contribution to the total columns at an impact parameter, $b=100$~kpc, for $f_{\rm shock}=1$. Each curve follows a single halo's trajectory. Left: \ovi contribution fractions; middle: \nv contribution fractions; right: \siiv contribution fractions. The gray rectangular frame roughly indicates the parameter space probed by \textit{COS}-Halos survey \citep{Werk_2016}.}
\label{fig:zvir-mvir-ovi}
\end{figure*}

In the bottom panel of Figure \ref{fig:Noviavg_vs_mstar} we display the fraction (percent) of the (halo-averaged) IGM column density from the (halo-averages) total column densities, for $f_{\rm shock}=1$. The dashed horizontal line marks a 50\% threshold. This figure indicates that on average, the IGM contribution never drops below $\sim40\%$. Over a significant range of stellar masses, the dominant contribution to the observed \ovi column seems to originate in the IGM rather than in the shock-heated halo.

This suggests that physical processes associated with the CGM may play smaller roles than often attributed to them,
when it comes to \ovi absorption.

\begin{figure*}
\centering
\includegraphics[width=1.0\textwidth, clip=True, trim={0cm 0cm 0cm 2cm}]{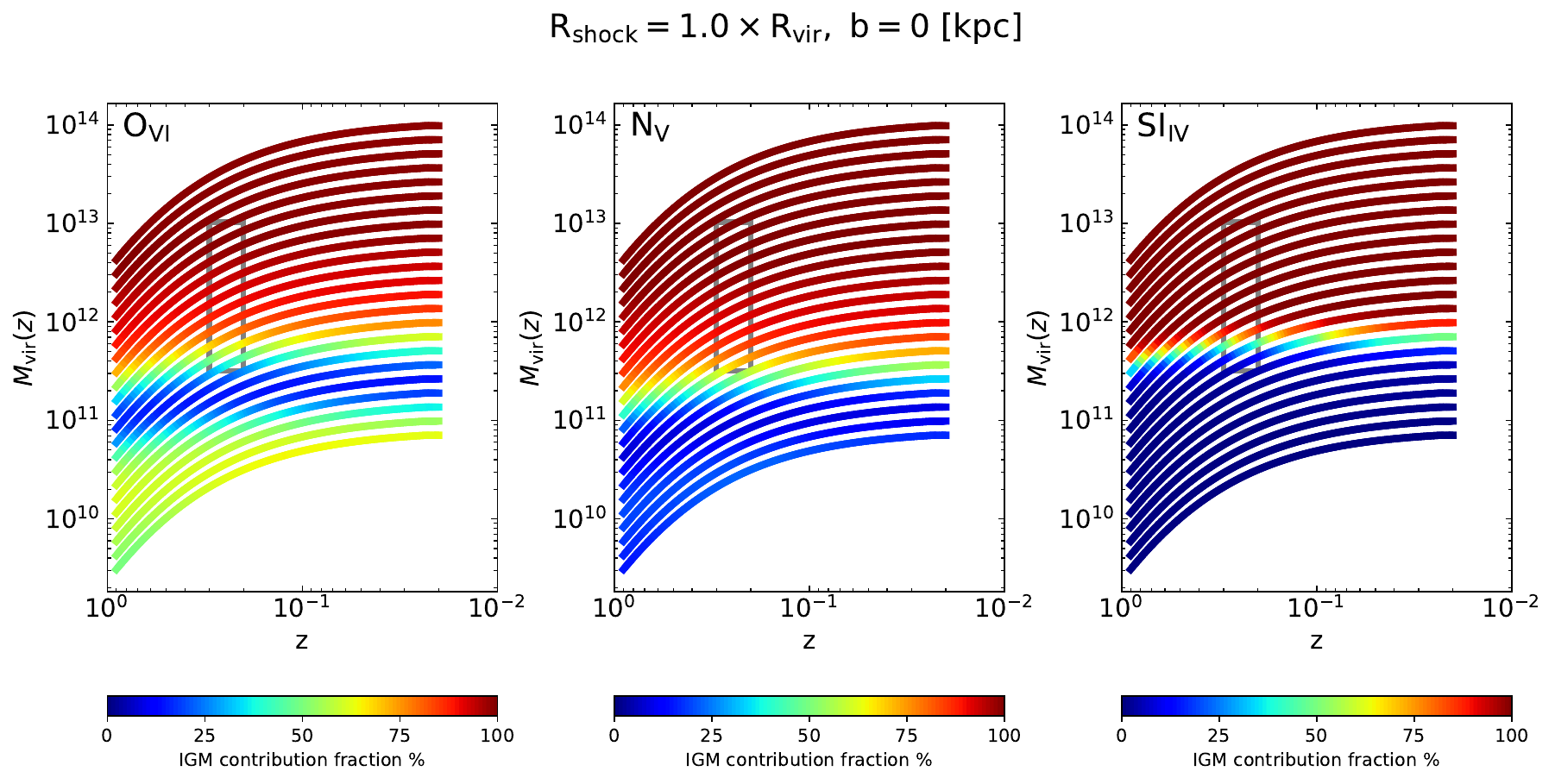}
\caption{Same as Figure \ref{fig:zvir-mvir-ovi}, but for an impact parameter $b=0~{\rm kpc}$. From left to right, we show the IGM contributions for \ovi, \nv, and \siiv column densities.}
\label{fig:zvir_mvir_ovi_no_b}
\end{figure*}

We now explore how these results depend on redshift, and whether they are specific to \ovi or may apply to additional warm-ion absorption signatures. 
We thus repeat the analysis for halos with masses in the range  $10^{11}-10^{14} ~M_{\odot}$, at redshifts in the range  $0.9<z<0.02$. 
In Figures \ref{fig:zvir-mvir-ovi} and  \ref{fig:zvir_mvir_ovi_no_b}, we display results assuming impact parameters,  $b=100$ and $0~{\rm kpc}$, respectively. 
Since observations at $b=0$ will realistically be dominated by the central galaxy, we only integrate starting at  $0.1R_{\rm vir}$, which we assume to be the galaxy radius. The columns for $b=0$ represent the maximum possible CGM contribution. We display results for \ovi, \nv, and \siiv.

Figure \ref{fig:zvir-mvir-ovi} follows the $\rm M_{vir}-z$ evolution of many halos. Each curve represents the evolution of a single halo and is color-coded by the fraction of IGM contribution to the total column density of \ovi (left panel), \nv (middle panel), and \siiv (right panel). The fractions are evaluated at an impact parameter $b=100$~kpc, assuming $f_{\rm shock}=1$. The figure shows that the IGM contribution is lowest for masses between $10^{11}-10^{11.6}~M_{\odot}$ and at low redshifts, $z\lesssim0.1$. Even then, the IGM contributes tens of percent of the total columns.

This trend may be readily explained by ion fraction distributions such as those displayed in Figure \ref{fig:zr-ions}. At high redshifts (low halo masses) the virial temperature is too low, and the halo gas density is too high, to allow efficient \ovi production either via photo-ionization or via collisional ionization. Later, intermediate mass halos have a virial temperature and densities ideal for efficiently producing \ovi within the CGM, especially in the inner parts of the halo. As the halos further grow and evolve, the virial temperature surpasses the optimal CPIE value for \ovi, thus diminishing CGM contribution. The relative CGM contribution therefore has an optimal mass/redshift regime. The same logic follows for \nv and \siiv.

In the IGM, however, the ion fraction profiles weakly depend on halo mass or redshift, ensuring a steady contribution to the total observed columns. 
This is due to the concerted action of two effects: the decline in density with cosmic evolution, and the decreasing intensity of the $UV$ background, which (for the redshift and mass range that we consider) result in a constant ionization parameter, to within a factor of $2$, in the nearby IGM.

Figure \ref{fig:zvir-mvir-ovi} shows that the IGM always significantly contributes to the total column density. 
For reference, we also mark a rough representation of the $M_{\rm vir}-z$ region (gray rectangle) probed by \textit{COS}-Halos survey \citep{Werk2013}.
In all three panels, the observed parameter space resides well within the region where the IGM contribution to the absorption-line columns is expected to be dominant.
This gives further indication that interpreting the observed columns as emanating from the CGM exclusively, may be inaccurate.

For completeness, we also show the IGM contribution fractions for lines of sight through the center of the halos (i.e. $b=0~{\rm kpc}$) in figure  \ref{fig:zvir_mvir_ovi_no_b}. The lower impact parameter naturally maximizes the halo's contribution to the ion columns. This is both due to the larger path length through the halo and because the inner parts of the halo produce \ovi more efficiently (see Figure \ref{fig:zr-ions}). The corresponding IGM fractions are therefore lower.
However, even in this case, the \textit{COS}-Halos survey parameter space mostly resides in regions where the IGM seems to dominate the observed columns.

\subsection{Simulated Absorption-Line Profiles}

\begin{figure}
\centering
\includegraphics[width=0.48\textwidth]{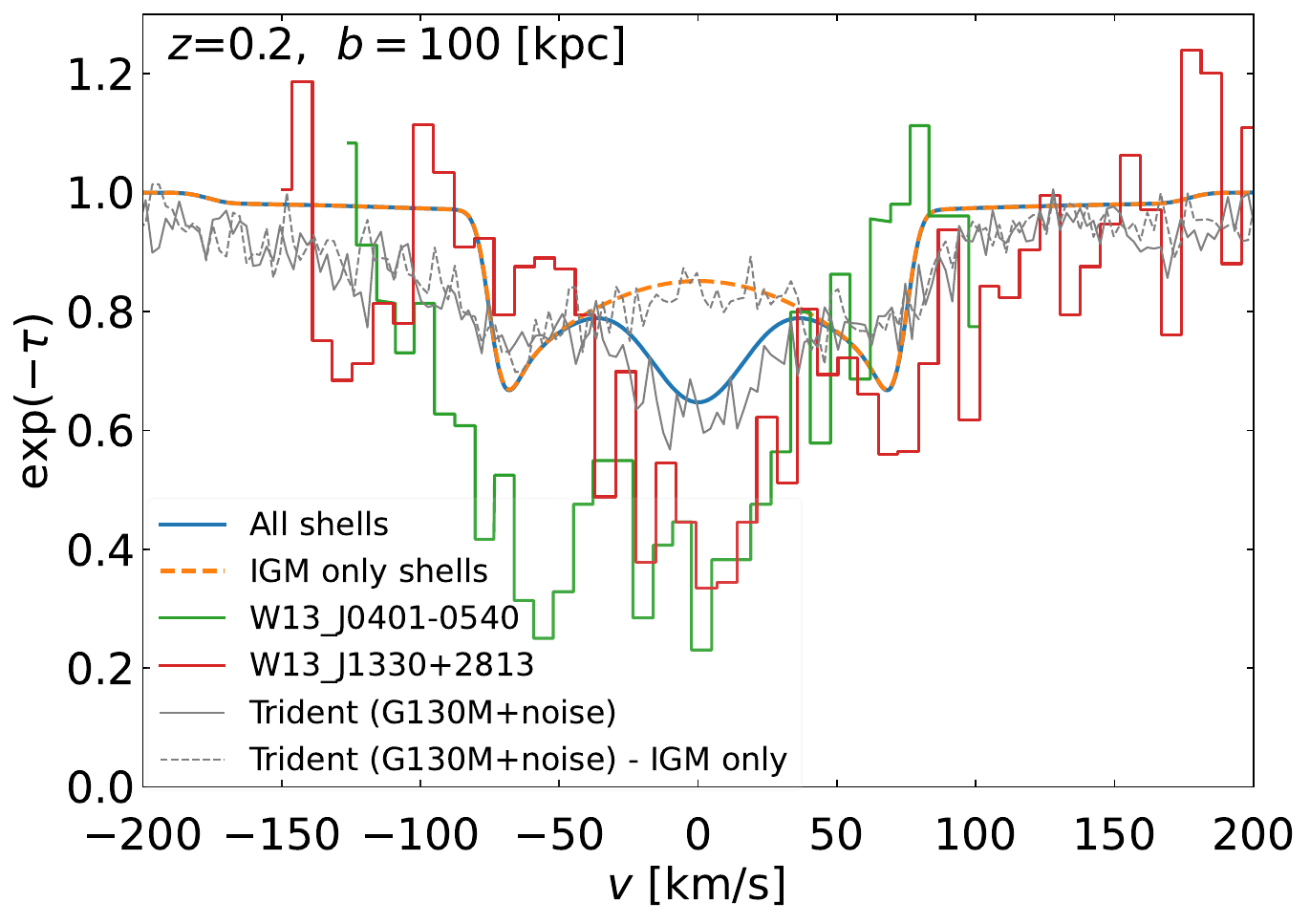}
\caption{Top: Simulated absorption line profiles for a $10^{12}~M_{\odot}$ halo at $\rm z=0.2$ assuming an impact parameter $b=100~{\rm kpc}$. The blue curve is the profile including all gas between $\rm 0.1R_{vir}-3~Mpc$. The dashed orange curve is for IGM only (namely, $\rm R_{shock}-3~Mpc$). The solid gray curve is a synthesized absorption line profile created with the \textit{Trident} Python package, which includes 
Milky-way foreground,
line-spread-function,
and Gaussian noise. The dashed gray curve is the \textit{Trident} result for the IGM only.}
\label{fig:absorption_line}
\end{figure}

Our simple $1D$ models suggest that significant fractions of \ovi and other warm-ions observed in CGM surveys may arise in the photoionized IGM surrounding the shocked halos.
We therefore explore whether the two components may be kinetically distinct. 
For this purpose, we compute the line profiles that are expected to arise due to the combined contributions of the virialized hot CGM and the warm, photoionized infalling IGM. As described in Section 2.5, we assume thermal broadening within each shell, as we integrate through the line-of-sight velocity profile for a given impact parameter (see Appx. \ref{Appendix: Absorption}). We verify our procedure by comparing to mock observations of \textit{COS} synthesized from our profiles by the \textsc{Trident} numerical package, as well as to observed line profiles from \cite{Werk2013}. 

Figure \ref{fig:absorption_line} shows a comparison between our results for the absorption line profile of the total gas (solid blue), and that of the IGM only (dashed orange). It also shows the absorption line produced by the \textsc{Trident} Python package for the same data, for the total gas (solid gray) and the IGM-only (dashed gray). The \textsc{Trident} profiles also include galactic foregrounds and noise to mimic HST-\textit{COS} observations. 

There is an apparent difference between the total and the IGM-only profiles near the line center, where the hot, virialized, low-velocity CGM gas is present. Aside from the line-center, there are two absorption-`dips', one from each side of the center. These are a result of the infalling IGM gas as well as the expanding gas beyond the turnaround radius.
The infall velocity is largest at $R_{\rm shock}$ where the gas is closest to the halo. This is also where the IGM density is maximal. The maximal infall velocity is set by the ratio between the halo mass and its radius ($v_{\rm max}^2 \approx G M_{\rm vir}/R_{\rm shock}$). 
A well-resolved kinematic structure of an absorption line may thus allow us to discern IGM from CGM contributions, as well as constrain the halo's physical characteristics.
For example, a kinematic structure composed of only two distinct absorption components may indicate that the IGM contribution dominates over the CGM columns.

While the kinematic difference between the total and IGM-only profiles seems sufficient for differentiation in our toy model, this "three-dip" structure will likely often not be distinct in observations of astrophysical halos, due to the complex morphology and kinematics of their multiphased CGM. 

To compare our idealized profiles with observational data, we display in figure \ref{fig:absorption_line}, as examples, data for two absorption line profiles extracted from \cite{Werk2013}: For J1330+2813 (red) and J0401-0540 (green).
Both lines exhibit stronger overall absorption than our model. The J1330+2813 line shows a dip structure similar to the one in our model, consisting of one central dip (at $\rm v\simeq0~km/s$), and two additional dips, one on each side of the line-center ($\rm v\simeq-130,75~km/s$). The J0401-0540 line shows a different structure, with two prominent dips ($\rm v\simeq0,-50~km/s$). Such structures may hint at additional contributions to the profile, other than the static halo, such as outflows, satellites, asymmetric accretion, etc.

\section{Conclusions \& Discussion}
\label{sec_conclusions}
In this work, we used a simple $1D$ toy model to
compute the cosmological collapse of dark matter halos and their baryonic content around small initial density perturbations. We computed the distributions and densities of the dark matter and gas as functions of time, from redshift $z=100$ up to the present day assuming a spherical collapse of the dark matter in an expanding EdS Universe. Using CLOUDY, we obtained the gas PIE temperatures due to the $UV$ background radiation, and the ion fractions of various ions both outside (PIE) and inside (forcing the temperature to $T_{\rm vir}$ when $T_{\rm vir}>T_{\rm PIE}$) the evolving halo. 

Our results indicate that a significant fraction of the column densities of warm ions (\ovi, \nv, and \siiv) observed near galaxies may originate from the inner IGM envelopes rather than the shocked CGM. The IGM envelopes are often neglected when interpreting such observations. 

We compute the IGM contribution fraction, as a function of the virial mass and redshift. We show that this fraction strongly depends on the halo's virial mass and on the impact parameter. 
For example, the minimal IGM contribution for \ovi columns is obtained for $M\simeq10^{11}-10^{11.6}~M_{\odot}$. Even then, the IGM contribution remains significant (with a minimum of$\sim15\%$ for an impact parameter of $0$~kpc).

For the mass range probed by most CGM surveys, (which is typically $\gtrsim 10^{11.5}~{\rm M}_\odot$), our model indicates that the IGM contribution to the total columns is significant and likely dominant.

The column densities computed with our models are in broad agreement with observational data, both within and outside the halos (see Figure 2), and also closely match the surface averaged values (Figure 3).

Our CGM model is quite basic and therefore fails to capture numerous intricate aspects of astrophysical halos, such as the morphological, thermal, and dynamical details of the multi-phased CGM or the radiative and dynamical properties of accretion shocks. Despite these limitations, it broadly matches the observed column densities both inside and outside the halo, as well as more complex estimates. We note, that the column densities predicted by our model are sensitive to the assumed density and metallicity profiles.
However, our main focus in this work has not been on a quantitative estimation of the observed column densities but rather on a qualitative evaluation of the IGM's contribution to the warm-ion absorption. We find our simplified model to be adequate for this specific purpose.

Finally, we composed synthetic absorption line profiles to test whether the CGM and IGM contributions to the total absorption may be discerned kinematically.
We demonstrated that a typical line profile is composed of a central (zero-velocity) wide component arising due to the hot CGM, and of two velocity-shifted features that emanate from the infalling IGM envelopes and expanding gas beyond turnaround, forming a "three-dip" structure. 
These synthetic spectra should, however, be taken with a grain of salt, since real-life halos do not have a perfectly static ($0$-velocity) CGM, but rather exhibit complex velocity structures: Galactic winds, stellar activity, and turbulence may all contribute to velocities of order hundreds km/s anywhere up to the virial radius. Thus, separating the CGM and IGM contributions in observed profiles will likely prove challenging (see Figure 6). Additional complications may arise due to asymmetric infall of the IGM since filaments, sheets, and satellite halos can also cause asymmetric line profiles, as often observed.

We conclude that the photoionized IGM envelopes of galactic halos may contribute significantly - and possibly dominantly -  to the observed warm-ion column densities in CGM surveys.
This contribution should be taken into account when attempting to gain insight into the physical processes in the CGM.
The inclusion of an IGM contribution may affect 
the inferred thermal properties and the conclusions regarding the role of non-thermal components and instabilities inside halos.

In a follow-up work, 
we intend to extend this study to include a more realistic, hydrodynamic spherical collapse model, which better represents the formation and structure of virial shocks. We will also include the influence of the $UV$ background and local ionizing sources, and take departures from ionization equilibrium into account. 
Our more realistic model may be more directly applicable to interpreting CGM observations.

\begin{acknowledgments}
This research was supported by the ISRAEL SCIENCE FOUNDATION (grant No. 2190/20).
\end{acknowledgments}

\software{Astropy \citep{astropy:2013,astropy:2018,astropy:2022},  
          CLOUDY \citep{ferland2017,ferland2023},
          YT analysis toolkit \citep{turkYtMulticodeAnalysis2011}
          Trident \citep{Hummels_2017}
          }

\bibliography{bibliography_file}{}
\bibliographystyle{aasjournal}

\appendix
\section{Kinematic \ion{O}{6} Absorption spectrum} \label{Appendix: Absorption}

In section 3.3, we present synthetic \ion{O}{6} absorption spectra for lines-of-sight passing through our computational box. Given an impact parameter $b$, the line-of-sight passes through all the shells with radial distances between $b$ and 3~Mpc. 

Each shell is characterized by its distance from the halo center $r_i$, \ion{O}{6} number density $n({\rm O_{\rm VI}})_i$, temperature $T_i$, and velocity relative to the halo centroid $v_i$. The projected line-of-sight velocity of a shell is then $v_{i, \scriptscriptstyle \parallel} = v_i \sqrt{1- (b/r_i)^2}$ (note, $b<r_i<3$~Mpc).  When computing the total absorption through the halo, each shell contributes twice $-$ on the near and far sides of the halo $-$ with opposite line-of-sight velocities, i.e., at $\pm v_{i, \scriptscriptstyle \parallel}$.

For the temperatures and optical depths characteristic of the CGM and IGM, the individual shell's line profiles are dominated by thermal broadening, and can therefore be well approximated by Gaussian profiles  \citep[e.g. eq. 6.38 from][expressed in terms of velocity]{Draine_book},
\begin{align*}
    \phi_i^{\pm}(v)\approx \frac{1}{\sqrt{\pi}\,b_i} \exp{ \left( -\frac{(v\pm v_{i,\scriptscriptstyle \parallel})^2}{b_i^2} \right)},
\end{align*}
which are centered about $\pm v_{i,\scriptscriptstyle \parallel}$ for the near and far sides of the halo.
Here $b_i=\sqrt{2k_BT_i / m }$ are the Doppler broadening parameters, and $\int\phi_i^{\pm}(v)dv=1$.
For \ion{O}{6} absorption, we set $m = m_{\rm O} \simeq 16 m_p$. With this line profile, the absorption cross-section is given by \citep[e.g.][eqn. 6.39]{Draine_book},
\begin{align*}
    \sigma^\pm_i(v)=\frac{\pi e^2}{m_ec}~f\,\lambda_0\,\phi^\pm_i(v) = \frac{\sqrt{\pi} e^2}{m_ec}~\frac{f\lambda_0}{b_i}\,e^{ -(v\pm v_{i,\scriptscriptstyle \parallel})^2/b_i^2},
\end{align*}
where $f$ is the oscillator strength for the transition.
For our \ion{O}{6} line at $\lambda_0 =1031.91 \rm$ \AA, the oscillator strength is $f({\rm O}_{\rm VI})=0.1376$ (\url{www.atomdb.org}).

The optical depth is then,
\begin{align*}
    \tau_v= \sum_i \left[ \sigma^+_i(v) + \sigma^-_i(v) \right] n({\rm O}_{\rm VI})_i~ds_i,
\end{align*}
where $ds_i$ is the path-length through shell $i$ along the line of sight, $ds_i = (r_{i+1}-r_{i})/\sqrt{1- (b/r_i)^2}$. 

Finally, the velocity-dependent intensity is given by,
\begin{align*}
    I(v)= I_0(v) e^{-\tau_v}.
\end{align*}
In creating Figure 6, we assumed a flat incoming spectrum $I_0(v)$.

\end{document}